\documentclass{article}
\usepackage{spconf,amsmath,amssymb,graphicx,booktabs,cite,xcolor}
\usepackage{tikz}
\usetikzlibrary{arrows.meta,fit,positioning}
\usepackage[hidelinks]{hyperref}

\definecolor{frozenfill}{HTML}{EEF1F5}
\definecolor{frozenedge}{HTML}{8C959F}
\definecolor{oursfill}{HTML}{E8F3EC}
\definecolor{oursedge}{HTML}{2F7D4F}
\definecolor{flowline}{HTML}{6E7781}
\definecolor{mutedtext}{HTML}{57606A}

\newcommand{\uh}{\emph{uh}}
\newcommand{\um}{\emph{um}}
\newcommand{\tagu}{\texttt{[uh]}}
\newcommand{\tagm}{\texttt{[um]}}
\newcommand{\anext}{a^{\mathrm{next}}_k}

\title{UMAI: UM Aware Inference with a Frozen Qwen3-ASR}
\name{Gyeongmin Kim}
\address{Department of Computer Science, Hanyang University, Seoul, South Korea\\
kdrkdrkdr@hanyang.ac.kr}

\begin{document}
\ninept
\maketitle

\begin{abstract}
Speech recognizers often leave filled pauses such as \uh{} and \um{} out of the transcript.
Qwen3-ASR-1.7B notices them but rarely writes them: at filler positions the filler token is usually ranked second or third and loses to the next word in the upper decoder layers.
We show that most of these fillers can be recovered from the decoder states and audio-encoder frames recorded during a single greedy decoding pass of the frozen recognizer.
For each gap between output words, UMAI (UM Aware Inference) combines them through two stages of linear discriminant analysis and isotonic calibration.
All statistics are estimated on the AMI meeting corpus without gradient descent, and no recognized word changes: UMAI only inserts \tagu{} and \tagm{} tags.
On the held-out AMI evaluation set, filler F1 rises from 12.2 to 82.6.
On the Nyra verbatim benchmark, where the operating threshold was chosen, it rises from 37.2 to 96.5.
Total processing time increases by 14\%.
\end{abstract}


\section{Introduction}
\label{sec:intro}

Verbatim transcripts keep what the speaker actually said, including fillers.
They can be used for clinical assessment of speech~\cite{wagner2024crisperwhisper}, natural text-to-speech data pipelines, and disfluency analysis~\cite{nyra_benchmark}.
Recognizers often drop them, however. The CrisperWhisper authors report that Whisper~\cite{radford2023whisper} eliminates many filler words~\cite{wagner2024crisperwhisper}.
The English set of the Nyra Verbatim Speech Benchmark~\cite{nyra_benchmark,nyra_data} has 3,072 filler tags (2,568 \texttt{[UH]} and 504 \texttt{[UM]}), and Qwen3-ASR-1.7B~\cite{qwen3asr} (hereafter Qwen3-ASR) writes 713 fillers on it.
Of these, 704 are correct (98.7\% precision), but recall is only 22.9\%.

Fine-tuning on verbatim transcripts can recover fillers~\cite{wagner2024crisperwhisper}, and prompting or biasing filler tokens could push decoding toward them.
All of these change what the recognizer decodes, so word choice, punctuation, and number formatting may change as well.
We study the problem under a stricter constraint: the recognized words stay exactly as they are, and the recognizer runs once.

We first checked whether Qwen3-ASR fails to notice these fillers or notices them and still does not write them.
We forced Qwen3-ASR to follow the reference text on 600 utterances sampled at random from the AMI~\cite{carletta2006ami} training meetings and looked at the 559 positions where the reference has a filler.
The filler token was mostly ranked second or third, and at 77\% of the positions it was among the top five; the top-ranked token was usually the word after the filler.
Table~\ref{tab:lens} applies the output layer directly to each decoder layer, as in the logit lens~\cite{nostalgebraist2020logit,belrose2023tuned}, at the 490 positions followed by a word.
In layers 17 to 19 the filler token is ranked ahead of the next word at most positions.
From layer 20 the next word overtakes it: the filler is ahead at 19\% of the positions by layer 23 and at 29\% at the last layer.
The information is in the model, and the filler loses to the next word at the output.

\begin{table}[t]
  \centering
  \caption{Median rank of the filler token and of the next word when the output layer is applied to each decoder layer, at 490 filler positions (the other 69 fillers end an utterance). Layers are counted from 1; 24 to 27 are omitted. Ahead: filler ranked above the next word (\%).}
  \label{tab:lens}
  \small
  \begin{tabular}{@{}lrrr@{}}
    \toprule
    Layer & Filler & Next word & Ahead \\
    \midrule
    1--16 & $10^2$--$10^4$ & $10^3$--$10^4$ & -- \\
    17 & 194 & 3,767 & 81 \\
    18 & 69 & 675 & 71 \\
    19 & 11 & 65 & 65 \\
    20--23 & about 20 & 9 to 1 & 45 to 19 \\
    28 (last) & 3 & 1 & 29 \\
    \bottomrule
  \end{tabular}
\end{table}

We ask whether the decoder states and audio-encoder frames of one decoding pass are enough to recover the fillers Qwen3-ASR leaves out.
The answer is largely yes. Our method, UMAI (UM Aware Inference), has two properties.
(1)~It reads values recorded during one greedy decoding pass and inserts \tagu{}/\tagm{} tags between output words; deleting the tags gives back Qwen3-ASR's text.
(2)~No parameter is trained by gradient descent: class means and shrunk covariances have closed forms, attention-head weights come from expectation--maximization (EM), and probability tables from isotonic regression.
The result is a 5.7\,MB file distributed separately from the recognizer, together with the code, at \url{https://github.com/kdrkdrkdr/umai}.

\section{Relation to Prior Work}
\label{sec:prior}

\textbf{Verbatim fine-tuning.} CrisperWhisper~\cite{wagner2024crisperwhisper} adjusts the tokenizer of Whisper~\cite{radford2023whisper} for more precise word-level timestamps and fine-tunes the model to produce more verbatim transcriptions.
UMAI keeps the original model and its text, and uses verbatim transcripts only to estimate statistics.

\textbf{Disfluency detection.} Many detectors work on transcripts, e.g., with a bidirectional LSTM~\cite{zayats2016disfluency} or a self-attentive parser~\cite{jamshidlou2020improving}; others detect dysfluencies in stuttered speech from wav2vec~2.0 features~\cite{bayerl2022detecting}.
UMAI instead reads the recognizer's own states, at positions given by its own words.

\textbf{Probing.} Linear classifier probes measure how suitable the features at every layer are for classification~\cite{alain2016understanding}, and self-supervised speech models have been analyzed layer by layer~\cite{pasad2021layerwise}.
We use a probe as a transcription tool whose decisions are scored against reference transcripts.
Because the probe is fitted on labeled data, part of what it detects may come from the fitting, so we do not claim that it measures only what Qwen3-ASR encodes.

\begin{figure*}[t]
  \centering
  \resizebox{\textwidth}{!}{%
  \begin{tikzpicture}[
      font=\small, >=Stealth, line width=0.6pt,
      box/.style={draw, rounded corners=2pt, align=center, inner sep=2.5pt, minimum height=0.95cm},
      frozen/.style={box, fill=frozenfill, draw=frozenedge},
      ours/.style={box, fill=oursfill, draw=oursedge},
      io/.style={box, fill=white, draw=frozenedge},
      flow/.style={->, draw=flowline},
      side/.style={->, draw=flowline, dashed},
    ]
    \node[io, minimum width=1.3cm]      (audio) at (0.65, 5.0) {Audio};
    \node[frozen, minimum width=3.4cm]     (enc)   at (3.65, 5.0) {\textbf{Audio encoder}\\24 layers + projector};
    \node[frozen, minimum width=4.1cm]     (dec)   at (8.55, 5.0) {\textbf{Text decoder}\\28 layers, greedy, run once};
    \node[io, minimum width=4.6cm]         (txt)   at (14.9, 5.0) {\textbf{Qwen3-ASR text (unchanged)}\\\textit{okay so we should go with \ldots}};
    \node[draw=frozenedge, dashed, rounded corners=3pt, fit=(enc)(dec), inner xsep=5pt, inner ysep=4pt,
          label={[anchor=south west, inner sep=1pt, text=mutedtext]north west:\textbf{Frozen Qwen3-ASR-1.7B}}] {};
    \draw[flow] (audio) -- (enc);
    \draw[flow] (enc) -- (dec);
    \draw[flow] (dec) -- (txt);

    \node[ours, minimum width=3.2cm] (post) at (3.65, 3.3) {\textbf{Frame posteriors} $p_t$\\two stacked 5-class LDAs};
    \node[ours, minimum width=2.6cm] (win)  at (7.25, 3.3) {\textbf{Audio window} $w_k(t)$\\alignment head};
    \node[ours, minimum width=3.2cm] (hs)   at (10.6, 3.3) {\textbf{Decoder states}\\steps $g$, $g{+}1$; 28 layers};
    \draw[flow] (enc.south) -- (post.north);
    \draw[flow] (win.north |- dec.south) -- (win.north);
    \draw[flow] (hs.north |- dec.south) -- (hs.north);

    \node[ours, minimum width=3.6cm] (aud)  at (5.45, 1.65) {\textbf{Audio features} $a_k$, $a^{\mathrm{next}}_k$\\pooled $p_t$, $\log p_t$ (20-d)};
    \node[ours, minimum width=3.2cm] (s1)   at (10.6, 1.65) {\textbf{Stage 1}\\per-layer LDA, 224-d};
    \node[ours, minimum width=4.0cm] (form) at (14.9, 1.65) {\textbf{\emph{uh} or \emph{um}}\\LDA + isotonic on $[\ell_k, f_k]$};
    \draw[flow] (post.south) -- (aud.north -| post.south);
    \draw[flow] (win.south) -- (aud.north -| win.south);
    \draw[flow] (hs) -- (s1);
    \draw[side] (hs.east) -| node[pos=0.25, above, text=mutedtext] {$\ell_k$: last layer at $g$} (form.north);

    \node[ours, minimum width=6.6cm] (s2)  at (8.0, 0.0) {\textbf{Stage 2}: 5-class LDA on $x_k$ (245-d) $\to$ isotonic $\to P_k$\\insert a tag if $P_k > \tau$};
    \node[io, minimum width=4.6cm]   (out) at (14.9, 0.0) {\textit{okay so [uh] we should go with \ldots}};
    \draw[flow] (aud.south) -- (s2.north -| aud.south);
    \draw[flow] (s1.south) -- (s2.north -| s1.south);
    \draw[flow] (s2) -- (out);
    \draw[flow] (form) -- (out);

    \node[frozen, minimum height=0.3cm, minimum width=0.3cm, inner sep=0pt] (lf) at (0.35, 2.0) {};
    \node[anchor=west, text=mutedtext] at (lf.east) {frozen};
    \node[ours, minimum height=0.3cm, minimum width=0.3cm, inner sep=0pt] (lo) at (0.35, 1.45) {};
    \node[anchor=west, text=mutedtext] at (lo.east) {UMAI};
  \end{tikzpicture}%
  }
  \caption{UMAI overview. Grey: the frozen Qwen3-ASR-1.7B, run once. Green: the UMAI read-out, estimated without gradient training.
  At each gap $k$ between output words, frame-class posteriors are pooled over two attention-derived audio windows and combined with the decoder states at steps $g$ (first token of the next word) and $g{+}1$.
  $f_k$ pools a projector-frame \uh/\um{} score over $w_k$.}
  \label{fig:overview}
\end{figure*}
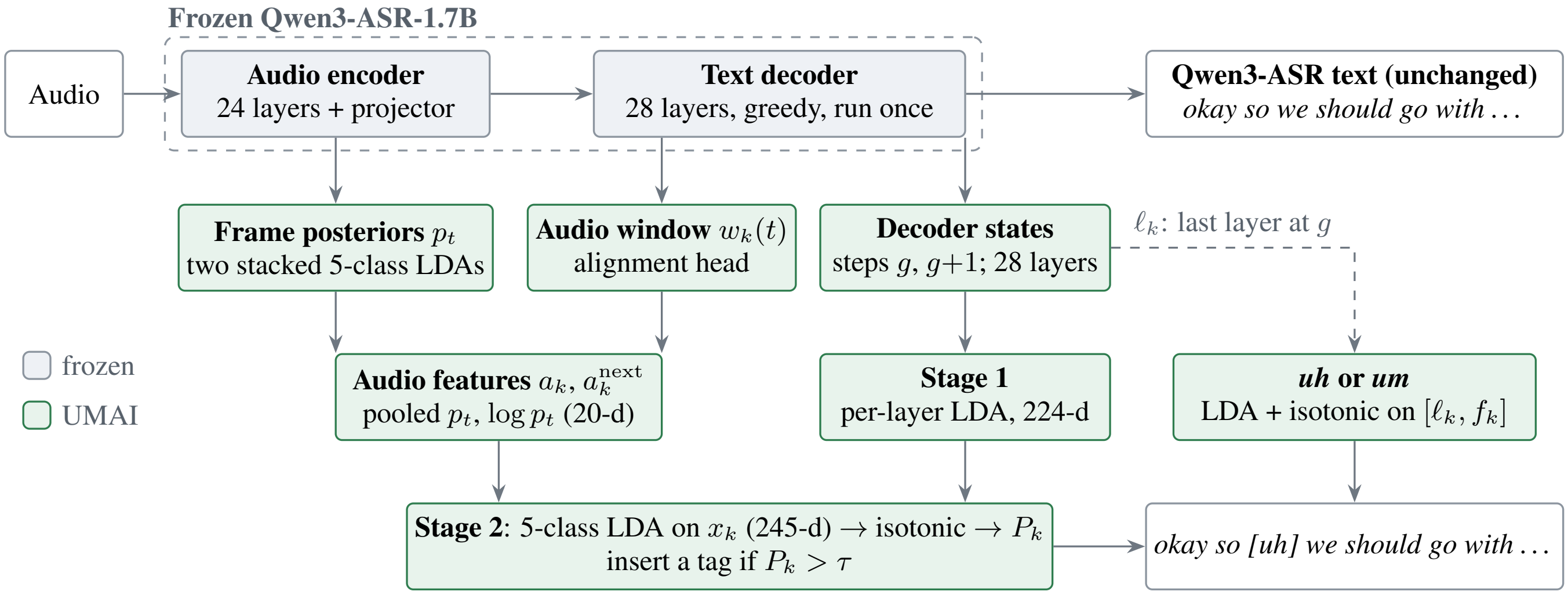

\section{Method}
\label{sec:method}

All statistics are estimated on a set of AMI utterances, the \emph{fitting data} (Section~\ref{ssec:fitting}). Fig.~\ref{fig:overview} gives an overview.

\subsection{Gaps}
\label{ssec:gaps}

Qwen3-ASR decodes greedily with the prompt of its official inference framework~\cite{qwen3asr}, with the language set to English, up to 260 tokens.
A \emph{word unit} is a whitespace-separated piece that contains an ASCII letter or digit.
The position before each word unit is a \emph{gap}, and so is the end of the utterance when decoding ends with the end-of-sequence (EOS) token.
No decision is made just before a filler that Qwen3-ASR wrote itself, and the backchannel \emph{uh huh} counts as one word.
For gap $k$, let $g$ be the step that emits the first token of the next word (the EOS step for the gap at the end of the utterance).
We record the outputs of all 28 decoder layers at steps $g$ and $g{+}1$, $h^{(l)}_g$ and $h^{(l)}_{g+1}$.
Step $g{+}1$ is recorded because much of the evidence lies outside step $g$ itself.
In an earlier single-pass classifier, correct insertions in cross-validation rose from 96 with only the top-ranked output candidates to 186 with internal states added, and to 348 with the states of the steps before and after added as well.
A gap with no step $g{+}1$ is an \emph{end gap}. Decisions are computed after decoding.

\subsection{Audio evidence}
\label{ssec:audio}

The vector of attention weights is often called the alignment~\cite{chorowski2015attention}.
We mix the attention of all 448 decoder heads (28 layers $\times$ 16 heads) into one alignment head, with weights $\pi$ fitted by EM~\cite{dempster1977maximum} to maximize $\sum_n \log \sum_i \pi_i M_{ni}$.
Here $M_{ni}$ is the attention mass, renormalized over audio frames, that head $i$ puts on the time span of word $n$ at the step that emits its first token.
The words are non-repeated reference words of two or more characters, sampled so that their durations (in 40\,ms bins) follow those of filler, cut-off, and repetition spans.
Only the sampled words that Qwen3-ASR recognized correctly are used.
The weight concentrates on one head (head 15 of layer 21, both counted from 1; $\pi_i = 0.91$); only three other heads receive more than 0.001.

Let $C_j(t)$ be the attention that the alignment head puts on frames $1,\dots,t$ at step $j$, and let $j_{\mathrm{prev}}$ be the step that emits the first token of the previous word. The audio window of gap $k$ is
\begin{equation}
  w_k(t) \propto C_{j_{\mathrm{prev}}}(t)\,\bigl(1 - C_g(t)\bigr)_+ ,
  \label{eq:window}
\end{equation}
normalized to sum to one, where $(x)_+ = \max(x, 0)$ and $C_{j_{\mathrm{prev}}} \equiv 1$ if no word precedes the gap.
It rises over the previous word and falls where the next word starts.

For every audio frame we compute posteriors $p_t$ over five classes (silence, word, filler, cut-off word, repetition) labeled from the reference word timings.
They come from two stacked LDAs: a five-class LDA on each of 25 sources (the 24 audio-encoder layers and the projector into the decoder), and a second LDA on the concatenated scores.
The audio feature of the gap is $a_k = [\sum_t w_k(t)\,p_t,\ \sum_t w_k(t)\log p_t]$ (10-d).
Pooling the same way over the window of the next word, $C_g(t)\,(1 - C_{g'}(t))_+$, gives $\anext$; here $g'$ is the next word-start or EOS step.
This feature is meant for cases where a stretched word onset sounds like a filler.
Features that cannot be computed (no audio frames, or an empty next-word window) are replaced by their means on the fitting data.

\subsection{Filler detection}
\label{ssec:filler}

Each gap in the fitting data gets one of five classes:
\textbf{O}, nothing is missing; \textbf{F}, the reference has \uh{} or \um{} here;
\textbf{E}, the reference has another word that Qwen3-ASR left out, such as a cut-off word or a repetition;
\textbf{U}, a neighboring word is misrecognized;
\textbf{Q}, the next word is a filler that Qwen3-ASR wrote.
Only F is the target; the other classes are meant to give the model separate explanations for an unusual decoder state.
UMAI makes no decision at Q gaps; keeping them as a class keeps their states out of the other classes.

\emph{Stage 1.} For each layer $l$ and step $j \in \{g, g{+}1\}$ we fit a five-class LDA with a covariance shared by the classes.
The covariance is shrunk by the oracle approximating shrinkage (OAS) estimator~\cite{chen2010shrinkage}, $\hat{\Sigma} = (1-\rho)S + \rho\,\tfrac{\operatorname{tr} S}{d} I$.
Here $S$ is the pooled within-class sample covariance, $d$ its dimension, and $\rho \in (0,1]$ has a closed form.
We keep the four log-posterior ratios against O, $28 \times 2 \times 4 = 224$ scores.

\emph{Stage 2.} The input $x_k$ (245-d) holds the 224 stage-1 scores, $a_k$, $\anext$, and a binary $b_k$ that is 1 if the piece just before the gap is a filler that Qwen3-ASR wrote.
A second five-class LDA gives $s_k = \log \frac{P(\mathrm{F} \mid x_k)}{1 - P(\mathrm{F} \mid x_k)}$.
End gaps lack the step-$(g{+}1)$ features and the next-word window, and use a separate stage-2 model (123 inputs) fitted on end gaps only.

\emph{Calibration.} Isotonic regression~\cite{zadrozny2002transforming} maps $s_k$ to a probability $P_k$, and UMAI inserts a filler when $P_k > \tau$ (default 0.75).
The fitting data are split into five folds by AMI meeting series. Stage-2 inputs and the isotonic fit use out-of-fold scores (ordinary stacking), and the stored models are then refitted on all data.

\subsection{Filler form (\uh{} vs.\ \um)}
\label{ssec:form}

We use Qwen3-ASR's own preference. Passing the last-layer state $h^{(28)}_g$ through Qwen3-ASR's final normalization and output layer gives the next-token distribution $P_{\mathrm{LM}}$, and
\begin{equation}
  \ell_k = \log \!\sum_{v \in \mathcal{V}_{\mathrm{um}}}\! P_{\mathrm{LM}}(v) - \log \!\sum_{v \in \mathcal{V}_{\mathrm{uh}}}\! P_{\mathrm{LM}}(v),
\end{equation}
over the 10 single-token spellings of \uh{} and \um. The softmax normalizer cancels, so only 10 output rows are computed.
On the audio side, an LDA on projector frames inside reference fillers gives a per-frame \um-versus-\uh{} log-likelihood ratio, and $f_k$ is its average over the gap window with weights $w_k(t)\,p_t(\text{filler})$.
A two-class LDA on $[\ell_k, f_k]$ followed by isotonic regression gives $R_k$, the probability of \um.
It is fitted on the 23,767 F gaps whose neighboring words match the reference.
UMAI writes \tagm{} if $R_k > 0.5$ and \tagu{} otherwise.

\section{Experimental Setup}
\label{sec:setup}

\subsection{Fitting data}
\label{ssec:fitting}

We use the headset audio, orthographic transcripts, and forced-alignment word timings of the AMI Meeting Corpus~\cite{carletta2006ami}.
The fitting data are the training meetings of the Full-corpus-ASR partition given on the AMI Corpus website, plus meeting IB4005 (533 utterances).
We leave out IS1003b, whose headset channels cannot be told apart by speaker, and three utterances that lie beyond the end of their audio file.
This leaves 66,900 utterances.
A filler is a reference word spelled \uh{} or \um.

Gap classes come from aligning Qwen3-ASR's words to the reference. The alignment labels a gap U whenever a neighbor is written differently, even when it stands for the same speech (\emph{gonna} vs.\ \emph{going to}).
An LLM reader (Gemma~4 31B~\cite{gemmateam2026gemma4}) therefore revisited each region that contained a U gap, or an F gap next to another missing word, when a reference filler lay within two words of it.
It saw both transcripts, the timed reference words, and each gap's alignment-derived class, but no audio.
It could also mark gaps where Qwen3-ASR wrote the filler as a different word; these gaps are excluded from fitting.
It changed 4,129 labels (2,641 U$\to$F, 1,475 U$\to$excluded, 7 F$\to$E, 5 F$\to$O, 1 F$\to$U).
On 1,164 gaps read twice by independent runs, agreement on F versus the rest was 98.9\% (Cohen's $\kappa = 0.974$), which measures consistency, not correctness.
The regions from an initial random sample of 19,874 utterances (1,252 of the 4,129 changes) were read under an earlier version of the instructions.
That version lacked one sentence stating that Qwen3-ASR's \emph{uh huh} is a backchannel word.
The double-read gaps come from the remaining 47,026 utterances.
The fitting set has 26,408 F, 19,595 E, 106,213 U, 655,977 O, and 1,922 Q gaps.
We use all non-O gaps (154,138) and as many sampled O gaps, weighted back to the original class frequencies.

\subsection{Test data}
\label{ssec:test}

\textbf{AMI eval}: the 16 evaluation meetings of the same partition, 7,490 utterances with 3,067 reference fillers.
No AMI eval speaker occurs in the fitting data, and no statistic, design choice, or threshold of UMAI was chosen on this set.
All scored outputs come from a single run.
(The timing test of Section~\ref{ssec:cost} additionally ran 155 of its utterances, 5 warm-up and 150 timed, without scoring.)
Design choices were compared on the AMI development meetings (8,665 utterances, not used for fitting; two of their speakers also speak in IB4005).
\textbf{Nyra}~\cite{nyra_benchmark,nyra_data}: 4,957 English utterances (all splits) with 3,072 reference fillers, based on DisfluencySpeech~\cite{wang2024disfluencyspeech}.
Nyra was used for one decision, the default $\tau = 0.75$, so its results at that threshold are optimistic; we report several thresholds.

\subsection{Metrics}
\label{ssec:metrics}

We use the scorer of the Nyra benchmark~\cite{nyra_benchmark}. An output filler, written by Qwen3-ASR or inserted by UMAI, is a true positive (TP) if aligned to a reference filler and a false positive (FP) otherwise.
A reference filler with no aligned output filler is a false negative (FN).
Filler F1 is $2\mathrm{TP}/(2\mathrm{TP}+\mathrm{FP}+\mathrm{FN})$; exact F1 also requires the form to match.
The scorer's exact count also takes as an FP any output word it reads as a sound-event label.
This adds four instances of \emph{noise} written by Qwen3-ASR on AMI eval and eight words on Nyra (\emph{noise} four times, \emph{breath} and \emph{laugh} twice each).
The disfluency hallucination rate is the number of output disfluency tokens aligned to no reference token, divided by the number of reference disfluency tokens.
The denominator is 4,488 on AMI eval (all fillers or cut-off words: 3,067 and 1,421) and 6,120 on Nyra.
The rate includes cut-offs and repetitions that Qwen3-ASR wrote.

We count only \uh{} and \um{} as fillers on both sides, as in the fitting data; the Nyra references contain no other forms.
Qwen3-ASR's \emph{hm}, \emph{hmm}, \emph{mm}, \emph{mhm}, \emph{ah}, and \emph{uh huh} therefore stay words instead of being normalized to filler tags.
Reference \emph{mm}, \emph{ah}, and \emph{hm} are also words (367 AMI eval tokens).
The output-side part of this rule was added after reading AMI eval errors (Section~\ref{ssec:errors}); it changes scoring, not outputs.
With the official normalization (keeping only \emph{uh huh} as a word), AMI eval has 3,434 reference fillers, and F1 is 16.5 for Qwen3-ASR alone and 78.1 for UMAI at $\tau = 0.75$.
Qwen3-ASR alone then has 322 TP and 142 FP, against 201 TP and 15 FP under our rule.
With only the reference-side part applied, the two scores are 12.3 and 79.4. On Nyra the output-side rule changes Qwen3-ASR alone from 706 TP and 10 FP to 704 and 9.

\section{Results}
\label{sec:results}

\begin{table}[t]
  \centering
  \caption{Filler tagging. Rows marked $\tau$ are UMAI. Ins.: tags inserted by UMAI. Ex.\ F1 requires the form; Hal.: disfluency hallucination rate (\%). The default $\tau = 0.75$ was chosen on Nyra.}
  \label{tab:main}
  \small
  \setlength{\tabcolsep}{2.6pt}
  \begin{tabular}{@{}llrrrrrrr@{}}
    \toprule
    Set & System & Ins. & TP & FP & FN & F1 & Ex.\ F1 & Hal. \\
    \midrule
    AMI & Qwen3-ASR & 0 & 201 & 15 & 2,866 & 12.2 & 12.1 & 0.65 \\
    eval & $\tau = 0.50$ & 2,557 & 2,473 & 300 & 594 & \textbf{84.7} & \textbf{82.1} & 4.34 \\
     & $\tau = 0.75$ & 2,216 & 2,270 & 162 & 797 & 82.6 & 80.0 & 2.30 \\
     & $\tau = 0.80$ & 2,144 & 2,211 & 149 & 856 & 81.5 & 79.0 & 2.12 \\
     & $\tau = 0.85$ & 2,005 & 2,096 & 125 & 971 & 79.3 & 76.9 & 1.78 \\
     & $\tau = 0.90$ & 1,912 & 2,010 & 118 & 1,057 & 77.4 & 75.1 & 1.65 \\
     & $\tau = 0.95$ & 1,194 & 1,358 & 52 & 1,709 & 60.7 & 58.8 & 1.05 \\
    \midrule
    Nyra & Qwen3-ASR & 0 & 704 & 9 & 2,368 & 37.2 & 36.6 & 0.41 \\
     & $\tau = 0.50$ & 2,403 & 2,967 & 149 & 105 & 95.9 & 94.3 & 1.73 \\
     & $\tau = 0.75$ & 2,268 & 2,920 & 61 & 152 & \textbf{96.5} & \textbf{94.9} & 0.72 \\
     & $\tau = 0.80$ & 2,247 & 2,906 & 54 & 166 & 96.4 & 94.7 & 0.67 \\
     & $\tau = 0.85$ & 2,208 & 2,879 & 42 & 193 & 96.1 & 94.5 & 0.56 \\
     & $\tau = 0.90$ & 2,176 & 2,852 & 37 & 220 & 95.7 & 94.1 & 0.49 \\
     & $\tau = 0.95$ & 1,836 & 2,527 & 22 & 545 & 89.9 & 88.3 & 0.44 \\
    \bottomrule
  \end{tabular}
\end{table}

\subsection{Comparison with the original model}
\label{ssec:main}

On AMI eval (Table~\ref{tab:main}), UMAI raises recall from 6.6\% to 74.0\% at $\tau = 0.75$ while precision stays about the same (93.1\% for Qwen3-ASR alone, 93.3\% for UMAI).
On Nyra, recall rises from 22.9\% to 95.1\% with 98.0\% precision.
With the form required (Ex.\ F1), F1 at $\tau = 0.75$ is 80.0 on AMI eval and 94.9 on Nyra, and the hallucination rate rises from 0.65\% to 2.30\% and from 0.41\% to 0.72\%.
Raising $\tau$ trades recall for fewer false positives.
Recall drops sharply between $\tau = 0.90$ and 0.95 (AMI eval 65.5\% to 44.3\%, Nyra 92.8\% to 82.3\%).
On AMI eval the best F1 is at $\tau = 0.50$, a lower threshold than the default chosen on Nyra.

UMAI scores much higher on Nyra than on AMI eval, although its statistics come from AMI.
Part of this may be optimistic bias, because $\tau$ was selected on Nyra, but Nyra F1 stays between 95.7 and 96.5 for $\tau$ from 0.50 to 0.90.
The sets also differ: Nyra is based on DisfluencySpeech, where one speaker reads conversational transcripts~\cite{wang2024disfluencyspeech}, while AMI eval is spontaneous multi-party meeting speech.
Qwen3-ASR alone also does much better on Nyra (22.9\% vs.\ 6.6\% recall). We did not separate these causes; AMI eval is our main estimate.

\begin{table}[t]
  \centering
  \caption{AMI eval against the relabeled reference, scored by gap. TP and FP include the fillers Qwen3-ASR wrote (201 TP, 15 FP).}
  \label{tab:relabel}
  \small
  \setlength{\tabcolsep}{4.5pt}
  \begin{tabular}{@{}lrrrrrr@{}}
    \toprule
    System & TP & FP & FN & Prec. & Rec. & F1 \\
    \midrule
    Qwen3-ASR & 201 & 15 & 2,547 & 93.1 & 7.3 & 13.6 \\
    $\tau = 0.50$ & 2,400 & 318 & 348 & 88.3 & 87.3 & \textbf{87.8} \\
    $\tau = 0.75$ & 2,217 & 171 & 531 & 92.8 & 80.7 & 86.3 \\
    $\tau = 0.80$ & 2,165 & 156 & 583 & 93.3 & 78.8 & 85.4 \\
    $\tau = 0.85$ & 2,056 & 132 & 692 & 94.0 & 74.8 & 83.3 \\
    $\tau = 0.90$ & 1,973 & 125 & 775 & 94.0 & 71.8 & 81.4 \\
    $\tau = 0.95$ & 1,345 & 56 & 1,403 & 96.0 & 48.9 & 64.8 \\
    \bottomrule
  \end{tabular}
\end{table}

\subsection{Relabeled AMI eval reference}
\label{ssec:relabel}

The scorer's word alignment has the same weakness as the alignment of Section~\ref{ssec:fitting}: a tag next to a misrecognized word may not be paired with the reference filler it stands for.
After the first results we therefore relabeled AMI eval with the same reader and region rule (758 regions, 1,480 gaps).
Regions were chosen from Qwen3-ASR's text and the reference alone, and the reader never saw UMAI's probabilities or tags.
All 206 aligned F gaps stayed F.
Of the 1,274 U gaps, 224 became F and 148 were excluded; the other 902 (679 read as O or E, 223 as U) keep the alignment label U and count as negatives.
On 156 gaps read twice, agreement on F versus the rest was 99.4\% ($\kappa = 0.982$).
A tag at an F gap is a TP, a tag at an O, E, or U gap is an FP, and an untagged F gap is an FN.
Excluded gaps are not scored, and neither are the tags UMAI places there (44 at $\tau = 0.75$).
All reference fillers between the same two Qwen3-ASR words form one F gap.
Fillers that Qwen3-ASR wrote are added as counted in Table~\ref{tab:main}, so the recall denominator is 2,748 (2,547 F gaps plus 201).

At $\tau = 0.75$, F1 is 86.3 (Table~\ref{tab:relabel}), 3.8 points above Table~\ref{tab:main} (86.33 vs.\ 82.56 before rounding).
Scoring by gap with the alignment-derived labels, before any reading, gives 83.45 (Qwen3-ASR alone 14.67).
Scoring by gap therefore accounts for about 0.9 of the 3.8 points, and the reading, including the 148 exclusions, for about 2.9.
Reading lowers Qwen3-ASR alone from 14.7 to 13.6.
For UMAI's tags alone, gap F1 at $\tau = 0.75$ is 82.3 and the average precision (AP) of $P_k$ over the scored gaps is 0.834 with the alignment-derived labels, and 85.4 and 0.900 after reading.
On the AMI development meetings, relabeled the same way under an earlier version of the instructions (165 gaps excluded), AP was 0.898.
Precision at $\tau = 0.75$ is similar in both scorings (93.3\% vs.\ 92.8\%), and most remaining errors are missed fillers (531 FN vs.\ 171 FP).

\subsection{Cost}
\label{ssec:cost}

UMAI adds no decoding pass.
From queries and audio keys recorded during decoding, it recomputes the attention of the 62 heads with nonzero weight (in 20 layers) at word-start and EOS steps.
It then scores the audio frames and applies a few small matrix products per gap.
We timed 150 consecutive utterances of one AMI eval speaker (501.7\,s of audio) on one GPU with batch size 1 and bfloat16.
The real-time factor rises from 0.139 to 0.158 (+14\%; mean 0.465\,s vs.\ 0.529\,s per utterance, median 0.187\,s vs.\ 0.228\,s).
Peak GPU memory rises from 4.08 to 4.19\,GiB.

\subsection{Where UMAI fails}
\label{ssec:errors}

Two LLM readers examined 60 false negatives and 60 false positives at $\tau = 0.75$ from the transcripts alone.
The errors were drawn from all 16 AMI eval meetings (three or four of each type per meeting), before the output-side scoring rule existed.
The readers agreed on 118 of 120 items.
Both disagreements were false negatives that one reader called ambiguous and the other a real miss; the counts below use one reader's labels.
Of the 60 false positives, 37 were words written by Qwen3-ASR, such as \emph{hmm} or \emph{ah}, that the official scorer counted as fillers (the reason for the rule; no longer counted).
Another 4 were tags at or next to a reference filler that the alignment did not pair with it, and 2 were ambiguous.
The remaining 17 were real errors, 13 of which put \uh{} next to the article \emph{a} or \emph{the}.
Of the 60 false negatives, 3 were ambiguous.
Of the 57 real misses, 30 were single fillers left untagged, and 12 were fillers that Qwen3-ASR wrote as another word (UMAI never replaces a word).
Another 8 were runs of two or three fillers that received one tag, and 7 were at the end of an utterance.
These small counts show which error types occur, not how often.
UMAI rarely tags a gap right after a filler that Qwen3-ASR wrote.
At $\tau = 0.75$ it tags 21 of 693 such gaps on Nyra (18 of them true positives) and 5 of 214 on AMI eval (4 true positives).

The same kind of read-out did not recover other disfluencies.
A cut-off word is already committed to a complete word (e.g., \emph{probably}, \emph{paper}) at its first token.
A repetition is collapsed by the decoder: the audio raises the repeated token from rank 13 to rank 2, but from layer 22 on the next word holds rank 1.
Of about 24,000 candidates only 511 were true repetitions, and at most 3 could be selected without a false insertion.
The decoder states carry almost no information about how many fillers a gap holds, so UMAI writes one tag per gap.

\section{Conclusion}
\label{sec:conclusion}

UMAI is a read-out built from linear discriminant analyses of values recorded during one greedy pass of a frozen Qwen3-ASR-1.7B.
It recovers most of the fillers the recognizer leaves out and changes no recognized word.
Filler F1 on AMI eval rises from 12.2 to 82.6 (13.6 to 86.3 against a gap-level relabeled reference) at 14\% extra processing time.
The read-out itself is a 5.7\,MB file.

\section{Compliance with Ethical Standards}
\label{sec:ethics}

This research study was conducted retrospectively using human subject data made available in open access by the AMI Meeting Corpus (CC BY 4.0) and the Nyra Disfluency Speech English dataset (Apache-2.0).
Ethical approval was not required, as confirmed by the licenses attached to the open access data.

\bibliographystyle{IEEEbib}
\bibliography{refs}

\end{document}